\documentclass[pre,amsmath,amssymb,aps,floatfix,nofootinbib]{revtex4}
\usepackage{dcolumn}
\usepackage{upgreek}
\usepackage{bm}
\usepackage{soul}
\usepackage{color}
\usepackage[dvips]{epsfig}
\usepackage{graphicx} 
\usepackage{amsmath} 
\usepackage{amssymb}
\usepackage{subfigure}
\usepackage{epstopdf}

\newcommand{\comment}[1]{}

\newcommand{\BEA}{\begin{eqnarray}}
\newcommand{\EEA}{\end{eqnarray}}

\newcommand{\bq}{\begin{equation}}
\newcommand{\eq}{\end{equation}}
\newcommand{\be}{\begin{eqnarray}}
\newcommand{\ee}{\end{eqnarray}}
\newcommand{\ba}{\begin{align}}
\newcommand{\ea}{\end{align}}

\renewcommand{\d}{{\rm d}}

\begin{document}

\title{Energy cost of dynamical stabilization: stored versus dissipated energy}

\author{A.E. Allahverdyan}
\affiliation{Alikhanian National Laboratory (Yerevan Physics Institute), 2 Alikhanian Brothers street, Yerevan 0036, Armenia}
\affiliation{Yerevan State University, 1 A. Manoogian street, Yerevan 0025, Armenia}

\author{E. Khalafyan}
\affiliation{Moscow Institute of Physics and Technology (State University), Moscow, Russia}

\begin{abstract} Dynamical stabilization processes (homeostasis) are
ubiquitous in nature, but energetic resources needed for their existence
were not studied systematically. Here we undertake such a study using
the famous model of Kapitza's pendulum, which attracted attention in
the context of classical and quantum control. 
This model is generalized, made autonomous, and we show that friction 
and stored energy stabilize the upper (normally unstable) state of the pendulum.
The upper state can be made asymptotically stable
and yet it does not cost any constant dissipation of energy, only a
transient energy dissipation is needed.  The asymptotic stability under
a single perturbation does not imply stability with respect to multiple
perturbations. For a range of pendulum-controller interactions, there is
also a regime where constant energy dissipation is needed for
stabilization. 
Several mechanisms are studied for the decay of dynamically stabilized states.
\end{abstract}

\maketitle

\section{Introduction}

Dynamical stabilization is an important concept in physics (particle
trapping, Floquet engineering) \cite{paul,cook,fish,polko}, control
theory (vibrational stabilization and robotics)
\cite{mech_1,mech_2,control,ieee}, biology (homeostasis)
\cite{review,billman,soodak,novo}, animal locomotion \cite{animal,animal_2}, and
population dynamics (polymorphism in time-dependent environment)
\cite{arm1,arm2}. The meaning of this concept is that certain relevant
parameters (concentrations, coordinates) are stabilized against external
perturbations by active and frequently self-regulating means. This is
achieved via specific engines or controllers, and no stability will
exist without their action. 

\comment{This is an important point that distinguishes homeostasis from standard
examples of physical stability that are achieved by passive means. For
example, relaxation of an oscillator to its lowest energy state under
friction is a passive process that does not need energy resources,
though an energy sink (friction) is necessary. }

Is there an energy cost for dynamic stabilization and how it is to be
estimated? This question is of obvious relevance for controlling
methods. A general explanation for homeostasis in biology is that it 
offers energetically cheaper realizations of physiological functions 
\cite{review}. This makes relevant to ask about its own energy costs.

In order to study the energy cost problem, we chose a simple but
non-trivial model that exhibits dynamical stabilization.  This is the
driven non-linear pendulum, whose upper (normally unstable state) can be
stabilized by a sufficiently fast external force thereby defying
gravity. Such models were first studied by Stephenon \cite{steph}, and
then by Kapitza \cite{kapitza,LL}; see \cite{butikov} for a review.
They still produce new physical results \cite{acheson,blackburn,fishman}
and have interesting applications 
\cite{paul,cook,fish,polko,mech_1,mech_2,animal,animal_2,control,ieee}.

Our first step will be to replace the external field with 
a controller degree of freedom in order to make the driven pendulum autonomous.
This ensures finite energies and accounts of all relevant
degrees of freedom; see section \ref{definition}. The autonomous
pendulum predicts the following two scenarios for dynamical
stabilization of the unstable state. Within the first scenario, the
state is asymptotically stable. There are two factors behind this strong
notion of stabilization: the energy stored in the controller that
ensures the needed effective potential, and the friction acting on the
pendulum. (Obviously, friction is necessary for asymptotic stability.)
There are no permanent energy costs here, i.e.  once the asymptotically
stable state is reached, the interaction with the controller is
automatically switched off. There is only a moderate transient
dissipation of energy during relaxation. The interaction
emerges on-line together with an external perturbation. 

However, the notion of asymptotic stability is not sufficient for
characterizing this scenario of dynamic stabilization.  Contrary to
passively stabilized systems, asymptotic stability does not guarantee
stability under a sequence of well-separated perturbations acting within
the attraction basin. For characterizing this more general notion of
stability we again need the concept of stored energy. 

The second scenario predicted by the model is realized when the
back-reaction from pendulum to controller is sizable. Here the the
asymptotic stability is replaced by a metastable stabilization that has
a finite (though possibly long) life-time, because the controller
steadily dissipates the stored energy for supporting the metastable
state. Once this energy is lower than a certain threshold, the
metastable state suddenly decays with dissipating away all the energy.
Therefore, dynamic stabilization within this scenario requires permanent energy costs.

Thus, the model provides conceptual tools and scenarios for addressing
the energy cost problem in more general (especially biological) dynamic
stabilization situations. 

This paper is organized as follows. The generalized Kapitza's model is
formulated in section \ref{definition}, and is approximately solved in
section \ref{solving}. We show that the asymptotic stability of the
inverted pendulum is determined (among other factors) by the energy
stored in the controller. This extends the stability criterion presented
in \cite{kapitza,LL,butikov}. In section \ref{solving3} we confirm the
analytic bound, and work out numerically two basic scenarios of dynamic
stability. Section \ref{ra} discusses a new scenario of noise-induced
metastability.  Our results are summarized in the last section, where we
also discuss possible biological implications of our research. 

\section{Pendulum and controller}
\label{definition}

\begin{figure}[t]
\includegraphics[width=3cm]{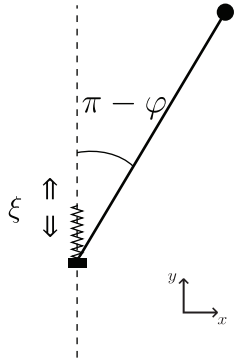}
\caption{A schematic representation of the inverted pendulum. The angle $\varphi$ is defined such that the
upper (normally unstable) position refers to $\varphi=\pi$. The end-point of the pendulum moves vertically
with coordinate $\xi$ that is subject to a harmonic potential; cf.~(\ref{3}).
}
\label{fig}
\end{figure} 

As a result of its numerous applications, Kapitza's pendulum can be
introduced in a variety of contexts.  For clarity, we shall introduce
its generalization within a mechanical picture; cf.~\cite{LL}. Consider
a pendulum moving on $(x,y)$ plane in a homogeneous gravity field $g$;
see Fig.~\ref{fig}.  The pendulum is a material point with mass $m$
fixed at one end of a rigid rod with length $l$.  Let the coordinates
and velocities of the mass be
\BEA
\label{1}
&& x = l \sin \varphi,  \quad   \dot x\equiv\frac{\d x}{\d t}  = l \cos \varphi \cdot \dot \varphi,\\
&& y = \xi_0 - l \cos \varphi, \quad \dot y\equiv\frac{\d y}{\d t} = \dot{\xi}_0 + l \sin \varphi \cdot \dot \varphi, 
\label{2}
\EEA
where $\xi_0$ refers to vertical (along $y$-axes) motion of the opposite end of the road. 
The  Lagrangian of the autonomous system with coordinates $(x,y,\xi)$ reads
\be
\label{3}
L = \frac{m}{2} [\dot x^2 + \dot y^2] - mgy + \frac{\mu}{2} \dot{\xi}_0 ^2 - \frac{k}{2} {\xi}_0^2,
\ee
where $\mu$ is the mass of $\xi_0$, and 
we assumed a harmonic potential $\frac{k}{2} {\xi}_0^2$ for $\xi_0$. Putting (\ref{1}, \ref{2}) into
(\ref{3}), denoting $\xi = \xi_0  + \frac{m g}{k}$, and dropping a constant term from Lagrangian we find
\BEA
\label{x2}
L = \frac{m l^2}{2} \dot \varphi ^2 + \frac{\mu + m}{2} \dot \xi ^2 
+  m l   \dot \xi \dot \varphi \sin \varphi + m g l \cos \varphi - \frac{k}{2} { \xi } ^2,
\EEA
which implies Lagrangian equations of motion $\frac{\d}{\d t}\frac{\partial L}{\partial\dot q}
=\frac{\partial L}{\partial q}$ with coordinates $q=(\varphi, \xi)$ and velocities 
$\dot q=(\dot\varphi, \dot\xi)$:
\BEA
\label{x3}
&&\ddot \varphi + \frac{g}{l} \sin \varphi 
= - \frac{1}{l} \sin \varphi \cdot \ddot \xi, \\ 
\label{x4}
&&\ddot \xi + \omega^2\xi = -\epsilon
\; [ \; \cos \varphi \cdot {\dot \varphi}^2 + \sin \varphi \cdot \ddot \varphi \; ], \\
&& \omega^2\equiv{k}/({\mu + m}), \qquad  \epsilon\equiv{l}/({1 + \frac{\mu}{m}}),
\label{x44}
\EEA
where $\omega$ is the frequency of $\xi$, while $\epsilon$ characterizes the back-reaction of $x$ on $\xi$.
Note from (\ref{3}, \ref{x4}) that whenever $\mu\gg m$, i.e. the controller $\xi$ is
much heavier than the pendulum, we can neglect the left-hand-side of
(\ref{x4}), and revert to the usual (non-autonomous) driven pendulum.
This, however, does not suffice for the full understanding of energy
costs, which arise due to the very back-reaction of $x$ on $\xi$. 

Eqs.~(\ref{x3}, \ref{x4}) are deduced from the time-independent Lagrangian (\ref{x2}); hence they are 
conservative and reversible. The conserved energy related to (\ref{x2}) reads:
\BEA
E=\dot\varphi\frac{\partial L}{\partial \dot\varphi}+\dot\xi\frac{\partial L}{\partial \dot\xi}-L=
\frac{m l^2}{2} \dot \varphi ^2 + \frac{\mu + m}{2} \dot \xi ^2 
+  m l   \dot \xi \dot \varphi \sin \varphi- m g l \cos \varphi + \frac{k}{2} { \xi } ^2.
\label{en}
\EEA

We shall now add a friction with parameter $\gamma>0$ to (\ref{x3}) writing it as
\BEA
\label{x5}
\ddot \varphi + \frac{g}{l} \sin \varphi + \gamma \dot \varphi
= - \frac{1}{l} \sin \varphi \cdot \ddot \xi.
\EEA
As seen below, this friction is a means of stabilizing the motion of $\varphi$. 
We do not add a friction to the controller degree of freedom $\xi$, since this will reach no 
constructive goal besides providing an additional channel for loosing energy. We shall 
also mostly neglect random noises acting on $\varphi$, i.e. we do not study Langevin equations. The 
influence of a random noise is discussed in section \ref{ra}.
 
Note that energy (\ref{en}) governed by (\ref{x4}, \ref{x5}) decays in time as it should:
\BEA
\label{ko}
\frac{\d E}{\d t}=-ml^2\gamma\dot\varphi^2\leq 0.
\EEA
As confirmed below, it is useful to separate the energy $E$ in (\ref{en}) into two contributions,
those descrbing the motion of $\xi$ and $\varphi$:
\BEA
E=E_\varphi+E_\xi,\qquad E_\xi= \frac{\mu + m}{2} \dot \xi ^2 + \frac{k}{2} { \xi } ^2.
\label{bull}
\EEA

\section{Solving the model via slow and fast variables}
\label{solving}

To solve non-linear (\ref{x4}, \ref{x5}), we shall apply both separation of time-scales and perturbation theory.
We assume that $\omega$ in (\ref{x4}) is a large parameter, i.e. $\xi$
oscillates fast. Next, we separate $\varphi$ as \cite{LL}:
\begin{equation} 
\label{eq:4}
\varphi = \Phi + \zeta, \qquad \zeta\ll\Phi
\end{equation}
where $\Phi$ is slow and $\zeta$ is fast, and also small compared with $\Phi$.
Then from (\ref{x4}, \ref{x5}, \ref{eq:4}) we get after expanding over $\zeta$ and keeping the first non-vanishing term only:
\BEA
\label{eq:5}
&& \ddot \Phi + \ddot \zeta + \frac{g}{l} \sin \Phi + \zeta \frac{g}{l} \cos \Phi 
+ \gamma \dot \Phi  +  \gamma \dot \zeta = - \frac{1}{l} \sin \Phi \cdot 
\ddot \xi - \frac{1}{l} \cos \Phi \cdot \zeta \ddot \xi,\\
&& \ddot \xi +\omega^2\xi = -\epsilon [(\cos \Phi - \zeta \sin \Phi)
(\dot \Phi + \dot \zeta)^2 + (\sin \Phi + \zeta \cos \Phi)(\ddot \Phi + \ddot \zeta)].
\label{eq:55}
\EEA
Assuming that $\gamma \gtrsim \frac{1}{\omega}$, and noting that 
for fast variables $\dot\zeta$, $\ddot\zeta$, $\dot\xi$, and $\ddot\xi$ 
are big, we can equalize fast and big components in (\ref{eq:5}, \ref{eq:55}):
\BEA
\label{eq:6}
&&\ddot \zeta +  \gamma \, \dot \zeta = - \frac{1}{l} \sin \Phi \cdot \ddot \xi,\\
\label{eq:66}
&&\ddot \xi + \omega^2 \xi = -\epsilon\sin \Phi \cdot \ddot \zeta,\\
&& \zeta (0) = 0,
\label{eq:77}
\EEA
where initial condition (\ref{eq:77}) is imposed without loss of generality; cf.~(\ref{eq:4}). 
In (\ref{eq:66}) we, in particular, neglected the factor $-\epsilon\cos \Phi \cdot \dot\zeta^2$,
because it is quadratic over $\zeta$. 
Note that (\ref{eq:6}, \ref{eq:66}) do not contain the contribution $\zeta \frac{g}{l} \cos \Phi $ coming from the potential
$-mgl\cos\varphi$ [cf.~(\ref{eq:5})], since this contribution is not sufficiently fast, i.e. (\ref{eq:6}, \ref{eq:66}) involve
only time-derivatives of $\zeta$. 

Eq.~(\ref{eq:6}) can be integrated over the time. A constant of integration should be put to zero, since $\zeta$
and $\dot\xi$ are oscillating in time with their time-average being zero. Hence the integration of (\ref{eq:6}) implies
\BEA
\label{ushi}
\dot\zeta(0)=-\frac{\sin\Phi}{l}\dot\xi(0).
\EEA
\comment{
The term $-\epsilon\cos \Phi \cdot \dot\zeta^2$ in (\ref{eq:66}) provides a non-zero average to $\xi$.}

Eqs.~(\ref{eq:6}, \ref{eq:66}) are linear over the unknown variables $\zeta$ and $\psi$, 
and they can be solved via the Laplace transform [see Appendix]. Now we can average (\ref{eq:5}) over fast oscillations:
\begin{equation}
\ddot \Phi +  \frac{g}{l} \sin \Phi + \gamma \dot \Phi + 
\frac{1}{l} \cos \Phi \cdot \overline{ \zeta (t) \, \ddot \xi (t)} = 0,
\label{bartok}
\end{equation}
where $\overline{...}$ means time-averaging over fast oscillations. 
The factor $\overline{ \zeta (t) \, \ddot \xi (t)}$
in (\ref{bartok}) is worked out in Appendix. It does contribute
to an effective (generally time-dependent) potential $\Pi(\Phi)$: 
\begin{equation}
\ddot \Phi + \gamma \dot \Phi =-\partial_{\Phi}\Pi(\Phi).
\label{ba}
\end{equation}
The form of this potential simplifies if we
assume that a slow variable $\beta\equiv\frac{\sin ^2 \Phi }{1 + \frac{\mu}{m}}$ holds $\beta\ll 1$ 
due to $\frac{\mu}{m}\gg 1$ (weak back-reaction), 
and take the first non-vanishing term over $\beta$; see Appendix. This approximation is only supported when 
the slow variable $\Phi$ relaxes to $\pi$ or to $0$. Eventually, we have a simplified form of the effective potential:
\begin{equation}
\label{ef}
\Pi = - \frac{g}{l} \cos \Phi - \frac{\omega ^2}{\gamma ^2 + \omega ^2} \cdot 
\frac{\dot\xi^2(0)+\omega^2\xi^2(0)}{8 l^2} \cdot \cos 2 \Phi.
\end{equation}
Now $\Phi = \pi$ is stable, i.e. $\partial_\Phi\Pi(\Phi)|_{\Phi=\pi}=0$ and $\partial^2_\Phi\Pi(\Phi)|_{\Phi=\pi}>0$, if:
\begin{equation}
\label{garni}
\frac{\dot\xi^2(0)+\omega^2\xi^2(0)}{2 g l} > 1 + \frac{\gamma ^2}{\omega ^2}.
\end{equation}
Note that larger values of $\omega$ expectedly increase the stability domain. However,
larger values of the friction constant $\gamma$ decrease it. Hence, the friction plays
a double role in this system, since the very relaxation of $\Phi(t)$ (e.g. $\Phi(t)\to\pi$) is achieved due
to friction \footnote{Note that Ref.~\cite{fishman} studied the inverted pendulum with friction and
deduced an effective potential that is akin to (\ref{ef}) (and even contains 
higher-order terms), but does not contain friction explicitly, since the latter was 
assumed to be small. }. 

Eqs.~(\ref{ba}, \ref{ef}) imply that when $\Phi=\pi$ is a stable rest
point, we get that the slow part $\Phi(t)$ of the angle variable
$\varphi(t)$ convergence due to the friction: $\Phi(t)\to\pi$, if
$\Phi(0)$ is in the attraction basin \footnote{Note that the fact 
of stabilizing the unstable rest-point $\varphi=\pi$ of the 
potential $-mgl\cos\varphi$ is due to the choice of this potential. Put differently, would 
we choose this potential as $-mgl\cos(\varphi-\varphi_0)$, then the effective potential will
still have the form  $\frac{\omega ^2}{\gamma ^2 + \omega ^2} \, 
\frac{\dot\xi^2(0)+\omega^2\xi^2(0)}{8 l^2} \, \cos 2 \Phi$; cf.~(\ref{ef}). Recall that 
(\ref{eq:6}, \ref{eq:66}) for fast variables|that eventually create the effective potential|do 
not contain the contribution coming from the potential. 
For the potential $-mgl\cos(\varphi-\varphi_0)$, $\Phi=\pi$ 
is (for a general $\varphi_0$) not a stable rest-point. 
} of $\Phi=\pi$. What happens then to the
fast part $\zeta(t)$ of $\varphi(t)$? This is a convoluted question that we clarify below numerically. Our
results in Appendix show that when the derivation
(\ref{eq:4}--\ref{ushi}) applies|i.e. both the time-scale separation and
the perturbation over $\beta$ hold|we get $\zeta(t)\to 0$ together with
$\Phi(t)\to\pi$; see (\ref{abo}). This is indeed observed numerically, as seen below.
However, there is also a regime that is not described by
(\ref{eq:4}--\ref{ushi}), where $\Phi(t)\to\pi$ for sufficiently long,
but finite times, and where $\zeta(t)$ stays non-zero; see below for
details. This regime is realized when the back-reaction parameter $\epsilon$ is sufficiently large. As we emphasized,
the above analytic derivations do not hold for this case. 
Eventually, the fact of large back-reaction leads to decaying of the $\Phi=\pi$
state, i.e. $\Phi=\pi$ turns out to be a metastable state. 

Finally, note that the left-hand-side of (\ref{garni}) is just the
initial dimensionless energy of $\xi$; cf.~(\ref{x44}, \ref{bull}). We
call this quantity the initial stored energy and remind again that
(\ref{garni}) was obtained for vanishing back-reaction $\epsilon\to 0$. 

\begin{figure}[!b]
\centering 
\subfigure[]{
    \includegraphics[width=0.35\columnwidth]{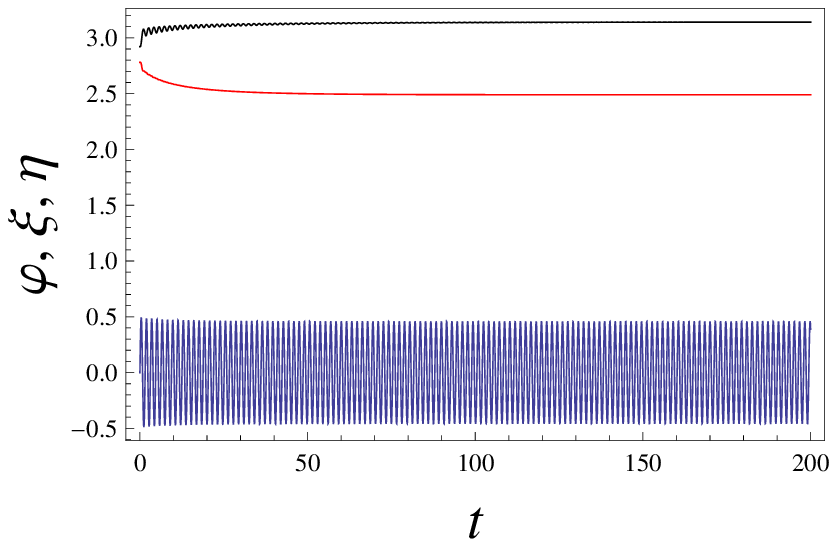}\label{fig1}
    } 
    \subfigure[]{
    \includegraphics[width=0.35\columnwidth]{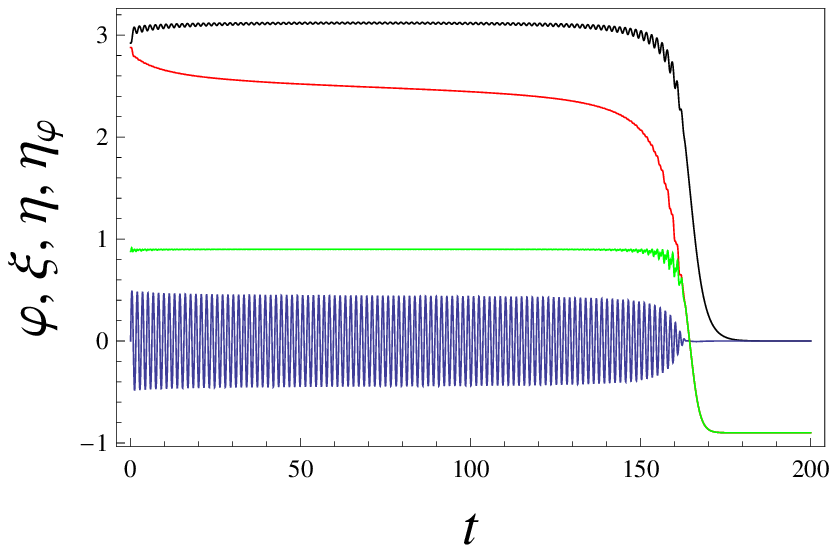}\label{fig2} 
    }
\caption{
(a) This figure shows the angle variable 
$\varphi$ (black, upper curve) and the controller $\xi$ (blue, oscillating curve) as numerical solutions of
(\ref{x4}, \ref{x5}) versus time $t$. The red curve (in the middle) is the
scaled energy $\eta=\frac{E}{\mu+m}$ that includes the stored energy $\frac{1}{2}[\dot\xi^2+\omega^2\xi^2]$; cf.~(\ref{en}, 
\ref{garni}). Parameters in (\ref{x4}, \ref{x5}) are: $\epsilon=0.8$,
$\dot\xi(0)=2$ ($\xi(0)=0$), $\varphi(0)=0.93 \pi$ ($\dot\varphi(0)=0$),
$\omega=4$, $\gamma=3$, $m=g=l=1$. \\ It is seen
that $\varphi$ quickly stabilizes at $\varphi=\pi$, which is normally an
unstable state.  The stabilization process takes a relatively small
amount of energy. After the stabilization $\varphi$ and $\xi$ decouple;
$\xi$ continues oscillating, and the scaled energy $\eta$ is then constant in time.
The stability condition (\ref{garni}) holds. (The difference between its LHS and RHS is $0.4375$.)\\
(b) The same parameters as in (a), but now the backreaction parameter
$\epsilon=0.9$ is slightly larger. We also display $\eta_{\varphi}=
\frac{E}{\mu+m}-\frac{1}{2}\dot \xi ^2 -\frac{\omega^2}{2} { \xi } ^2$ (green curve, third from top), where 
$\eta_{\varphi}$ is the (scaled) energy related to the angle variable 
only; cf.~(\ref{en}). Now there is rather long ($t\sim
150$) period of metastability, accompanied by a slow dissipation of
energy. After this the stability is lost: $\varphi$ quickly relaxes to
the minimum $\varphi=0$ of the potential, $\xi$ looses all its
energy and eventually stops moving (i.e. $\xi(t)\to 0$). The whole stored
energy is dissipated away. \\
The physical reason of scenario in (b) is that fast oscillations around 
$\varphi=\pi$ do not disappear, i.e. they persist in the metastable state, 
continuously dissipate energy [cf.~(\ref{ko})], and once the initial energy decreases sufficiently,
the $\varphi$ and $\xi$ (relatively) suddenly move to the global energy minimima 
$\varphi=\dot\varphi=\xi=\dot\xi=0$.
We emphasize that it is the energy $\frac{1}{2}\dot \xi ^2 +\frac{\omega^2}{2} { \xi } ^2$ 
stored in $\xi$ that decays in time. To confirm this, (b) also shows the scaled energy $\zeta_\varphi$ related to $\varphi$.
It stays constant in time for the whole metastability period; see the green curve. 
A similar scenario happens, when the initial condition $\varphi(0)$ is
out of the attraction basin of $\varphi=\pi$. However, here $\varphi(t)$ never reaches $\pi$. 
}
\end{figure}

\section{Scenarios of (de)stabilization for the inverted pendulum} 
\label{solving3}

\subsection{Asymptotic stability and stability with respect to several perturbations}
\label{olaf}

Fig.~\ref{fig1} shows solution of (\ref{x4}, \ref{x5}) when condition
(\ref{garni}) holds. It is seen that indeed the state $\varphi=\pi$ gets
asymptotically stabilized, $\varphi(t)\to\pi$ at least when $\Phi(0)$ is
sufficiently close to $\pi$, i.e. when $\varphi(0)$ is in the attraction
basin of $\pi$. This relaxation is accompanied by the energy
dissipation.  The decaying (dissipating) quantity here is mostly the
stored energy $\frac{1}{2 g l}[\dot\xi^2(t)+\omega^2\xi^2(t)]$, i.e. the
dimensionless energy related to $E_\xi$ in (\ref{bull}). Once $\varphi$
approaches $\pi$ sufficiently close, the coupling between $\xi$ and
$\varphi$ is switched off; recall the discussion after (\ref{garni}). This means that the stored
energy is not anymore dissipated and stays constant for subsequent
times; see Fig.~\ref{fig1}. Thus, we confirm that the originally unstable fixed point of the pendulum
can be made asymptotically stable without permanent energy costs, but with a transient energy dissipation
only. 

For parameters of Fig.~\ref{fig1} the point $\varphi=\pi$ is
asymptotically stable with a well-defined attraction basin; i.e.  it is
stable with respect to a single perturbation, which refers to the
initial state $\varphi(0)=0.93\pi$, $\dot\varphi(0)=0$, and specific
initial state $(\xi(0), \dot \xi(0))$ of $\xi$; see Fig.~\ref{fig1}. 
It is necessary to generalize the notion of
asymptotic stability, because it is unrealistic to consider only a
single perturbation. Let us assume that after $\varphi(t)$ relaxed to
$\pi$, we apply at some random time $\tau$ (which is larger than the relaxation
time) yet another (second) perturbation
$\varphi=\pi\to\varphi(0)=\varphi(\tau)$ within the same attraction
basin, i.e. $\varphi(0)=0.93$ for parameters of Fig.~\ref{fig2}. The
initial state of $\xi$ is then reset as $(\xi(\tau), \dot \xi(\tau))$.
Hence we now run the dynamics anew with initial states
$\varphi(0)=0.93\pi$, $\dot\varphi(0)=0$, $\xi(\tau)$ and $\dot
\xi(\tau)$. 

It appears that for parameters of Fig.~\ref{fig1}, $\varphi=\pi$ is
unstable after the second perturbation.  The issue here is that the
back-reaction $\epsilon$ is sufficiently large, hence the second
perturbation alters the controlling degree of freedom $\xi(t)$ 
and also dries out its stored energy, which
already decreased after the first perturbation. If for parameters of
Fig.~\ref{fig1} we decrease $\epsilon$ from $0.8$ to $0.1$, $\varphi$
becomes stable to many well-separated (in the above sense) perturbations
coming at random times. One reason for this is that the stored energy
decrease within one perturbation is smaller. Another reason, which 
numerically clearly differs from the first one, is that 
the motion of $\xi$ becomes more stable with respect to resetting 
the initial conditions of $\varphi$ during the perturbation. 

\subsection{Random noise}
\label{ra}

Above we assumed multiple, strong and well-separated (in time)
perturbation.  Another way of implementing multiple perturbations is to
include an external noise in (\ref{x5}). Let us add to the right-hand-side
of (\ref{x5}) a white, Gaussian random noise:
\BEA
\label{gau}
\sigma f(t), \qquad \langle f(t)\rangle=0, \quad \langle f(t)f(t')\rangle=\delta(t-t'),
\EEA
where $\sigma$ is the noise intensity. The modified (\ref{x5}) becomes 
then the Langevin equation (for $\varphi(t)$). In contrast to strong and
well-separated (in time) perturbations, (\ref{gau}) allows uncorrelated,
densely located perturbations that are (most probably) weak if $\sigma$ is small. 

\begin{figure}[!b]
\centering
\subfigure[]{
    \includegraphics[width=0.35\columnwidth]{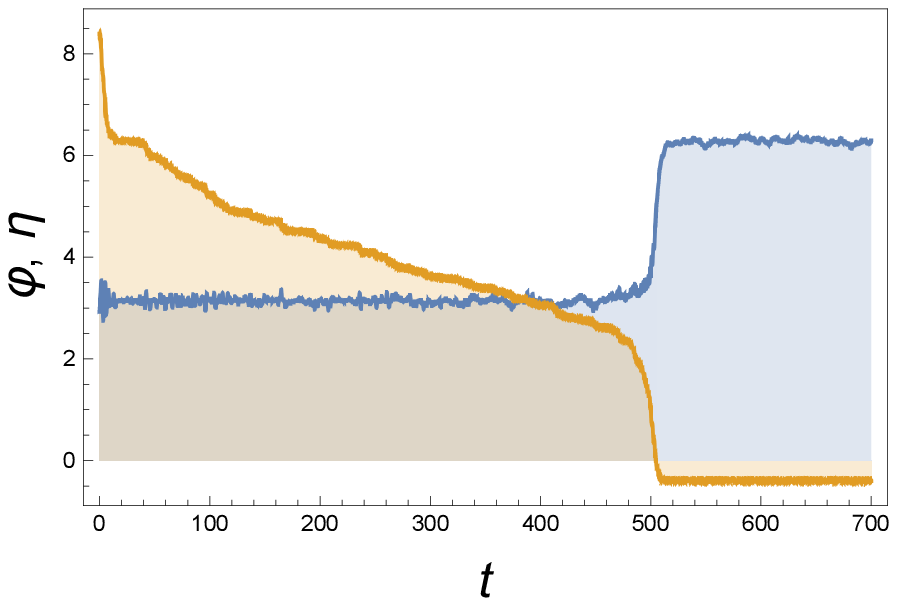}\label{fig5}
    } 
    \subfigure[]{
    \includegraphics[width=0.35\columnwidth]{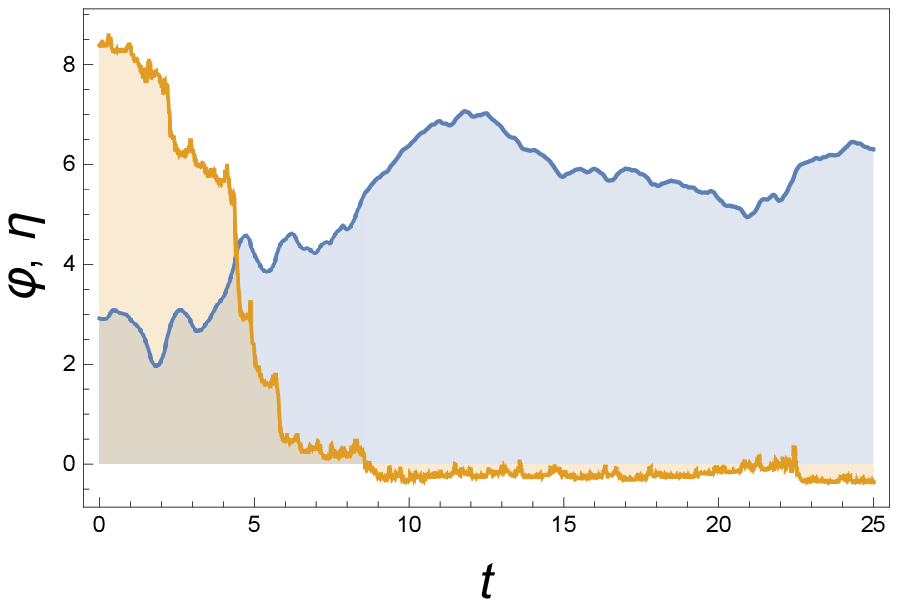}\label{fig6} 
    }
\caption{(a) The angle $\varphi$ (blue, piece-wise constant curve) as the numerical solution of
(\ref{x4}, \ref{x5}) versus time $t$, where to the RHS of (\ref{x5}) we 
added an addtive white noise $f(t)$ with intensity $\sigma=0.1$; see (\ref{gau}). 
Here $\eta=\frac{E}{\mu+m}$ (orange, decaying curve) is the
scaled energy; cf.~(\ref{en}). Parameters in (\ref{x4}, \ref{x5}) are: $\epsilon=0.4$,
$\dot\xi(0)=4$ ($\xi(0)=0$), $\varphi(0)=0.93 \pi$ ($\dot\varphi(0)=0$),
$\omega=4$, $\gamma=3$, $m=g=l=1$.\\ It is seen that $\phi$ is stabilized around $\pi$ in a metastable state whose life-time is 
$\simeq 500$. During this life-time the energy $\eta$ slowly decays, till it is below some critical value, and then the
metastable state suddenly decays. \\
(b) The same as in (a), but for a stronger noise $\sigma=1.25$. 
Metastable decay is smeared, but still clearly visible.
}
\label{fig56}
\end{figure}

Now Fig.~\ref{fig5} shows that a weak-random noise monotonously dries
out the stored energy, and hence the stability of $\varphi=\pi$ is lost
after a sufficiently long time. Fig.~\ref{fig6} shows that the same
scenario hold for a stronger noise, though in somewhat blurred form and
for a shorter life-time of the metastable state. For parameters of
Fig.~\ref{fig5} and $\sigma=0$, $\varphi=\pi$ is asymptotically stable
with respect to $10-11$ strong, well-separated perturbations. 

This metastability under noise differs from the well-known 
noise-induced escape from a local energy minimum to a deeper minimum.
There a sufficiently strong random kick brings the system to
the attraction basin of the deeper minimum. In contrast, here the
attraction basin is not changed (and no strong kicks are to be waited 
for). Rather the stored energy needed for maintaining the stability 
is drained out and the very local minimum state is destroyed. Hence, 
what we described here constitutes a new type of noise-induced 
metastability that deserves further study.

\subsection{Metastability}

For parameters of Fig.~\ref{fig1}, if the back-reaction $\epsilon$ is larger than $0.8$
another interesting scenario takes place: the notion of asymptotic stability with respect 
a single perturbation is lost and is replaced by metastability; see Fig.~\ref{fig2}. 

Now small oscillations of $\varphi(t)$ around $\Phi=\pi$ persist and do not
decay in time; cf.~(\ref{eq:4}). According to (\ref{ko}), these
oscillations slowly drain out the initial stored energy $\frac{1}{2 g l}
[\dot\xi^2(0)+\omega^2\xi^2(0)]$, and when it gets sufficiently low, the metastable
state $\Phi=\pi$ suddenly decays to the global energy minimum
$\varphi=\dot\varphi=\xi=\dot\xi=0$; see Fig.~\ref{fig2}. During this
sudden decay the whole stored energy is dissipated away. Note that in
the metastability time-window the energy $E_\varphi$ related to
$\varphi$ stays constant; see (\ref{bull}) and Fig.~\ref{fig2}. 

The transition between regimes in Fig.~\ref{fig1} and Fig.~\ref{fig2},
i.e. between the truly stable and the metastable state takes place for a
critical value of the backreaction parameter $\epsilon_c$. For
parameters of Fig.~\ref{fig1} and \ref{fig2} we have
$\epsilon_{c}\approx 0.86075$. We checked numerically that the life-time
of the metastable state can be very large for $\epsilon$ approaching
$\epsilon_c$ from above. Hence we conjecture that this time can be
arbitrary large for $\epsilon\to\epsilon_c+0$. 

We emphasize that the transition between regimes in Fig.~\ref{fig1} and
Fig.~\ref{fig2} is described for fixed initial conditions of $\varphi$:
$\varphi(0)=0.93 \pi$ ($\dot\varphi(0)=0$), i.e. for a fixed attraction
basin of the stabilized state. If these initial conditions are changed
by making $\varphi(0)$ closer to the stability point $\pi$ (i.e. the
attraction basin is shrunk), then the transition from the stable to
metastable regime takes place at a larger value of $\epsilon$, or does
not take place at all. Likewise, if the attraction basin is enlarged,
the transition taken place for smaller $\epsilon$. For example, if
$\varphi(0)=0.92\pi$ ($\dot\varphi(0)=0$), then the transition takes
place at $\epsilon_c\simeq 0.6536$, while for $\varphi(0)=0.935\pi$
($\dot\varphi(0)=0$), the solution is stable for all $\epsilon<1$, which
is the physical range of $\epsilon$ for parameters of Fig.~\ref{fig1}
and Fig.~\ref{fig2}; cf.~(\ref{x44}). Note that the stabilization with
the largest attraction basin demands vanishing values of $\epsilon$. In
particular, for parameters of Fig.~\ref{fig1}, no stabilization of the
$\varphi=\pi$ state occurs for $|\varphi(0)|<0.64105\pi$
($\dot\varphi(0)=0$), while for $|\varphi(0)|\gtrsim0.64105\pi$ the
stabilization demands $\epsilon\to 0$.

Note that the metastability in Figs.~\ref{fig5} and \ref{fig6} is
different from that in Fig.~\ref{fig2}, because in the latter case the
metastability is due to a strong back-reaction, and not permanently
acting perturbations.

\section{Summary and Discussion}

The purpose of this work is to understand energy costs of dynamical
stabilization (homeostasis): a process that stabilizes an unstable state
due to an active controlling process. To this end, one needs plausible
models with a well-established history of physical
\cite{paul,cook,fish,polko,steph,kapitza,LL,butikov,acheson,blackburn,fishman}
and control-theoretic \cite{mech_1,mech_2,animal,animal_2,control,ieee}
applications. Here we studied the inverted (Kapitza's) pendulum model,
where the upper (normally unstable) state is stabilized by a fast motion
of a controlling degrees of freedom. Usually, this degree of freedom is
replaced by an external field. But here we modeled it explicitly,
because we want to study an autonomous system with a full control of
energy and its dissipation. Our main results are summarized as follows. 

The unstable state of the pendulum can be asymptotically stabilized|with 
a finite attraction basin|without
permanent energy dissipation, because the controller-pendulum interaction is
automatically switched off once the pendulum is stabilized. Here there
is only a transient dissipation of a small amount of energy related to
the stabilization.  This regime is reached when both the backreaction of
the pendulum to the controller is sufficiently small and the controller
oscillates sufficiently fast, i.e. it does have a sizable stored energy. 

However, the notion of asymptotic stability is not sufficient: we need
to study stability with respect to multiple perturbations. We
implemented two scenarios for multiple perturbations: strong, widely
separated in time perturbations, and a weak white noise acting on the
pendulum. An asymptotically stable state may not be
stable with respect to several perturbations. The latter type of
stability is achieved only if the back-reaction to the controller is
small. In the second scenario, a weak white noise leads to a
noise-driven decay of a metastable state. This decay differs in several
ways from the usual noise-driven escape. There is a need for further, 
more systematic research on this effect.

When the back-reaction is larger, the very notion of asymptotic
stability is lost and is replaced by metastability. Now the
stabilization is temporary (metastable), because small oscillations
around the stabilized state do not decay. They dry out the stored energy
of the controller, and once it is lower than some threshold the
metastable state decays. In this case, we do get that a constant
dissipation rate is needed for supporting the metastable state. 

Note that the energy cost problem was actively studied for the case of
adaptation \cite{gorban}. Here the stability is required with respect to
external changes of intensive variables; e.g. temperature, chemical
potential {\it etc} \cite{sartori_nature,wang,sartori_prl,prl,gorban}.
Adaptation should be distinguished from the proper dynamical
stabilization (homeostasis) \cite{c_gorban}.  Adaptation is about
stability of intensive variables (e.g.  temperature). It relates to
structural changes in the system, while the homeostasis need not. Hence,
adaptation is thermodynamically restricted by the Le Chatelier-Braun
principle, whereas homeostasis is not \cite{wang,c_gorban,gilmore}. The
existing approaches to energy cost of adaptation show that the energy is
to be dissipated continuously if an adaptive state is maintained
\cite{sartori_nature,wang,sartori_prl,prl}.  In that sense adaptation is
similar to proof-reading and motor transport, biological processes that
are essentially non-equilibrium and demand constant dissipation of
energy; see \cite{jsp} for a review. 
The question of relating dynamical stabilization and adaptation more 
directly remains open; so far, the two concepts have been quite separate.

Conceptual tools gained from the physical model should be useful for 
studying the energy cost of biological examples of dynamical 
stabilization (homeostasis)
\cite{review,billman,soodak,novo}. Biological and physiological
discussions imply that homeostasis is needed for controlling (and
providing advantages for) metabolic processes in organisms
\cite{review,billman,soodak,novo}. This makes necessary to ask about the
proper energy costs of the homeostasis itself. Such costs can be
substantial, e.g.  humming birds (colibri)|for
which energy saving is crucial|fall at night into a torpor state that is
different from the normal sleep. In this state several homeostatic
mechanisms|including internal energy regulation|are ceased. Thereby
birds are able to save a substantial amount of energy: up $\simeq 60 \%$
of the normal usage \cite{colibri}. 

Since realistic models of homeostasis are derived from systems biology
\cite{review,gorban}, they frequently lack a physical form that allows
us to ask questions about energy balance, let alone its dissipation.
Nonetheless, some analogies can be drawn and might prove useful in
future research. We saw that the energy stored in the controlling degree
of freedom can be the main resource of homeostasis. In that
respect it is similar to the energy stored in the living organism, one
of major concepts in biological thermodynamics
\cite{bauer,mcclare,jaynes,blumenfeld,arshav,debt1,debt2}. It also
relates to adaptation energy introduced in physiology \cite{gorban};
cf.~\cite{c_gorban} for a critical discussion. The stored energy is
phenomenologically employed in Dynamic Energy Budget Theory (DEBT)
\cite{debt1,debt2} and applied for estimating metabolic flows of
concrete organisms. 

This concept is not yet well-formalized, but some of its qualitative
features are known. The stored energy is not the usual free energy,
since the latter is present in equilibrium as well. From the viewpoint
of a modern thermodynamics, the notion of the stored energy resembles
the energy kept at certain negative temperature, because it is capable
of doing work in a cyclic process \cite{balian}. Now for the inverted
pendulum the stored energy is mechanical (not chemical) and relates to
an oscillating degree of freedom, but it is also capable of doing a
cyclic work. The inverted pendulum does demonstrates how stored energy
relates to stabilizing unstable states. 

\comment{
Yet another scenario when the stabilization is metastable and needs a 
constant dissipation of energy is when the degree of freedom to be stabilized is
subject to a weak, external random noise. This scenario is expected, since it 
amounts to permanently acting small perturbations, each one dissipating 
its portion of stored energy, as in the above transient regime. }

\comment{
This result can be interesting from at least two viewpoints. First,
minimizing the stored energy can be relevant in control theory, where
the inverted pendulum widely applies \cite{animal,animal_2,control,ieee}. Second,
it is known that living organisms do not receive the whole their energy
from the food: the energy coming from external signals is much smaller,
but still relevant in sustaining the living state \cite{jo}. We believe
that the above result of reducing the stored energy and partially
replacing it with the energy coming from the perturbation itself can
qualify as a simple initial model, where the trade-off between the
stored and received energy can be explored and studied in more detail. 
}

\begin{acknowledgments}
This work was supported by SCS of Armenia, grants No. 20TTAT-QTa003 and No. 21AG-1C038. 
A.E.A. was partially supported by a research grant from the Yervant Terzian 
Armenian National Science and Education Fund (ANSEF) based in New York, USA. 
\end{acknowledgments}

\appendix

\section{Solution of Eqs.~(\ref{eq:6}, \ref{eq:66})}

Define from (\ref{x44}) 
\BEA
\label{beta}
\beta \equiv \frac{\epsilon \, \sin ^2 \Phi}{l}=\frac{\sin ^2 \Phi }{1 + \frac{\mu}{m}}<1,
\EEA
and solve (\ref{eq:6}, \ref{eq:66}) via the Laplace transform as:
\BEA
\label{a1}
\hat\xi (s)=\frac{(1-\beta)a(s)}{(s^2+\omega^2)(s+\gamma)-\beta s^3},\qquad
\hat\zeta (s)=-\frac{\sin\Phi}{l}\, 
\frac{(1-\beta)b(s)}{(s^2+\omega^2)(s+\gamma)-\beta s^3},\\
a(s)=(s+\gamma)\dot\xi(0)+s(s+\frac{\gamma}{1-\beta})\xi(0),\qquad
b(s)=s\dot\xi(0)-\xi(0)\frac{\omega^2}{1-\beta}
\EEA
where in addition to initial condition (\ref{eq:77}) we also employed 
(\ref{ushi}). 

The inverse Laplace transform taken from (\ref{a1}) reads
\BEA
\label{abel}
\xi(t)=\left[
\frac{e^{s_1t}a(s_1)}{(s_2-s_1)(s_3-s_1)}+\frac{e^{s_2t}a(s_2)}{(s_1-s_2)(s_3-s_2)}+
\frac{e^{s_3t}a(s_3)}{(s_1-s_3)(s_2-s_3)}\right],\\
\zeta(t)=-\frac{\sin\Phi}{l}\left[
\frac{e^{s_1t}b(s_1)}{(s_2-s_1)(s_3-s_1)}+\frac{e^{s_2t}b(s_2)}{(s_1-s_2)(s_3-s_2)}+
\frac{e^{s_3t}b(s_3)}{(s_1-s_3)(s_2-s_3)}\right],
\label{abo}
\EEA
where $s_1$, $s_2$ and $s_3$ solve 
\BEA
s^3+\frac{\gamma s^2}{1-\beta}+\frac{\omega^2 s}{1-\beta}+\frac{\omega^2\gamma}{1-\beta}=(s-s_1)(s-s_2)(s-s_3)=0.
\label{kub}
\EEA
Without loss of generality we take the following parametrization:
\BEA
\label{topo}
&& s_1=-\Gamma,\qquad s_2=-\widetilde{\gamma}+i\widetilde{\omega},\qquad s_3=-\widetilde{\gamma}-i\widetilde{\omega},\\
&& \Gamma>0,\qquad \widetilde{\gamma}\geq 0.
\EEA
Now we calculate $\overline{ \zeta (t) \, \ddot \xi (t)}$. Hence when taking the product $\zeta (t) \, \ddot \xi (t)$
all oscillating terms with frequency $\widetilde{\omega}$ are to be neglected:
\BEA
\overline{ \zeta (t) \, \ddot \xi (t)} = -\frac{\sin\Phi}{l}\left(
\frac{s_1^2a(s_1)b(s_1)\, e^{-2\Gamma t}}{[(\widetilde{\gamma}- \Gamma)^2 + \widetilde{\omega}^2]^2}+
\frac{ (s_2^2a(s_2)b(s_3)+s_3^2a(s_3)b(s_2) )\, e^{-2\widetilde{\gamma}t}  }{4\widetilde{\omega}^2
[(\widetilde{\gamma}- \Gamma)^2 + \widetilde{\omega}^2]}
\right).
\label{vle}
\EEA

\comment{
This is the particular case of (\ref{vle}) for $\xi(0)=0$.
\BEA
\overline{ \zeta (t) \, \ddot \xi (t)} = -\frac{\sin\Phi\, \dot\xi^2(0)}{l}\left(
\frac{\Gamma^3(\Gamma-\gamma)\, e^{-2\Gamma t}}{[(\widetilde{\gamma}- \Gamma)^2 + \widetilde{\omega}^2]^2}-
\frac{ (\widetilde{\gamma}^2+ \widetilde{\omega}^2)(
\widetilde{\omega}^2 + \gamma\widetilde{\gamma}-\widetilde{\gamma}^2 )\, e^{-2\widetilde{\gamma}t}  }{2\widetilde{\omega}^2
[(\widetilde{\gamma}- \Gamma)^2 + \widetilde{\omega}^2]}
\right).
\label{vle2}
\EEA
}

If in (\ref{beta}), $\beta\to 0$ due to $\mu/m\gg 1$, we can keep in
(\ref{kub}) only the simplest non-zero order putting there $\beta=0$.
This approximation is supported if the slow variable $\Phi$ tends to
$\pi$, i.e. $\beta$ gets additional smallness. Now (\ref{kub}) will 
read $(s+\gamma)(s^2+\omega^2)=0$, i.e. we get via (\ref{topo}):
\BEA
\Gamma=\gamma,\quad \widetilde{\gamma}=0,\qquad \widetilde{\omega}=\omega. 
\label{mort}
\EEA
This means that the term $\propto e^{-2\Gamma t}$ in (\ref{vle}) can be neglected, and we have
\BEA
\label{avan}
\overline{ \zeta (t) \, \ddot \xi (t)} = \frac{\sin\Phi\,[ \dot\xi^2(0)+\omega^2\xi^2(0)]}{l}
\,\frac{\omega^2}{2(\gamma^2+\omega^2)}.
\EEA
Eq.~(\ref{avan}) leads us to (\ref{bartok}, \ref{ef}).

\end{document}